\documentclass[sigconf,natbib,screen=true]{acmart}

\usepackage{graphicx}
\usepackage{booktabs} 
\usepackage{numprint}
\npthousandsep{,}
\npdecimalsign{.}
\usepackage[inline]{enumitem}
\usepackage{balance}
\usepackage[skip=0pt]{caption}
\usepackage{multirow}
\usepackage{siunitx}
\usepackage{acronym}
\usepackage{subcaption}

\usepackage{etoolbox,siunitx}
\robustify\bfseries

\copyrightyear{2020}
\acmYear{2020}
\setcopyright{rightsretained}
\acmConference[SIGIR eCom '20]{The 2020 SIGIR Workshop On eCommerce}{July 30, 2020}{Xi'an, China}
\acmBooktitle{Proceedings of the 2020 SIGIR Workshop on eCommerce (SIGIR eCom'20), July 30, 2020, Xi'an, China}
\acmPrice{}

\definecolor{darkgreen}{rgb}{0.0, 0.2, 0.13}

\newcommand{\header}[1]{\vspace{1mm}\noindent\textbf{#1}.}
\newcommand{\headerl}[1]{\vspace{1mm}\noindent\textit{#1}.}

\AtBeginDocument{%
  \providecommand\BibTeX{{%
    \normalfont B\kern-0.5em{\scshape i\kern-0.25em b}\kern-0.8em\TeX}}}

\acrodef{IR}{information retrieval}
\acrodef{SERP}{search engine result page}

\newcommand{\justabit}{$\, \, \, \, \,$}

\author{%
Fatemeh~Sarvi$^1$ \justabit
Nikos~Voskarides$^2$ \justabit
Lois~Mooiman$^3$ \justabit
Sebastian~Schelter$^{2,4}$ \justabit
Maarten~de~Rijke$^{2,4}$
}
\affiliation{%
  \institution{
  $^1$AIRLab, University of Amsterdam \justabit
  $^2$University of Amsterdam \justabit
  $^3$Bol.com \justabit
  $^4$Ahold Delhaize 
  }
}  
\email{[f.sarvi, n.voskarides, s.schelter, m.derijke]@uva.nl, lmooiman@bol.com}

\def\authors{Fatemeh Sarvi, Nikos Voskarides, Lois Mooiman, 
Sebastian Schelter and Maarten de Rijke}

\setcopyright{rightsretained}
\acmConference[SIGIR eCom'20]{ACM SIGIR Workshop on eCommerce}{July 30, 2020}{Virtual Event, China}
\acmYear{2020}
\copyrightyear{2020}
\makeatletter
\renewcommand\@formatdoi[1]{\ignorespaces}
\makeatother
\acmISBN{}

\begin{document}
\sisetup{detect-weight=true,detect-inline-weight=math}

\title[A Comparison of Supervised Learning to Match Methods for Product Search]{A Comparison of Supervised Learning to Match\\ Methods for Product Search}

\renewcommand{\shortauthors}{Sarvi, Voskarides, Mooiman, Schelter and De Rijke}
 

\begin{abstract}
The vocabulary gap is a core challenge in \ac{IR}. In e-commerce applications like product search, the vocabulary gap is reported to be a bigger challenge than in more traditional application areas in \ac{IR}, such as news search or web search.
As recent learning to match methods have made important advances in bridging the vocabulary gap for these traditional \ac{IR} areas, we investigate their potential in the context of product search. 

In this paper we provide insights into using recent learning to match methods for product search. 
We compare both effectiveness and efficiency of these methods in a product search setting and analyze their performance on two product search datasets, with $\sim$50,000 queries each.
One is an open dataset made available as part of a community benchmark activity at CIKM 2016.
The other is a proprietary query log obtained from a European e-commerce platform.
This comparison is conducted towards a better understanding of trade-offs in choosing a preferred model for this task. 
We find that
\begin{enumerate*}
	\item models that have been specifically designed for short text matching, like MV-LSTM and DRMMTKS, are consistently among the top three methods in all experiments; however, taking efficiency and accuracy into account at the same time, ARC-I is the preferred model for real world use cases; and
	\item the performance from a state-of-the-art BERT-based model is mediocre, which we attribute to the fact that the text BERT is pre-trained on is very different from the text we have in product search.
\end{enumerate*}	
We also provide insights into factors that can influence model behavior for different types of query, such as the length of retrieved list, and query complexity, and discuss the implications of our findings for e-commerce practitioners, with respect to choosing a well performing method.
\end{abstract}



\maketitle

\acresetall


\section{Introduction}
\label{sec:intro}

Online shopping is gaining in popularity~\citep{tsagkias-2020-challenges}. 
E-commerce platforms offer rich choices in each of (often) many categories to the point that finding the desired article(s) can be impossible without an adequate search engine.
In this context, an effective product search engine benefits not just the users but also suppliers.

\header{Implications of the vocabulary gap in product search.} The vocabulary mismatch between query and document poses a critical challenge in search~\citep{li-2014-semantic}. 
The vocabulary gap occurs when documents and queries, represented as a bag-of-words, use different terms to describe the same concepts. While BM25~\citep{robertson2009probabilistic} continues to be a reliable work horse in practical search engines, there is a growing collection of neural learning to match methods aimed specifically at overcoming the vocabulary gap.  These methods go beyond lexical matching by representing queries and documents in finite-dimensional vector spaces and learning their degree of similarity in this space~\citep{mitra2017neural,onal-neural-2018}. In product search, the vocabulary gap may be a larger problem than in other \ac{IR} domains~\citep{vangysel-learning-2016}. Product titles and queries tend to be short, and titles are not necessarily well-structured sentences but consist of phrases or simple combinations of keywords. 

\header{Semantic matching} While product search leverages a wide range of ranking features~\citep{tsagkias-2020-challenges}, features that do not rely on popularity or past interaction behavior, are deemed to be important as well. Semantic matching is one of the most important techniques to improve the ranking in product search~\citep{vangysel-learning-2016,tsagkias-2020-challenges}.
Several semantic matching methods have already been applied in the area of product search to generate latent representations for queries and product descriptions~\citep{zhang2019improving, zhang2019neural,tsagkias-2020-challenges}. Surprisingly, despite recent advances in supervised learning to match methods (see Section~\ref{section:relatedwork} for an overview), relatively little is known about the performance of these methods in the setting of product search.

\header{Our experimental study} We fill this gap by conducting a systematic comparison of 12 supervised learning to match methods on the task of product search. We compare the ranking performance of these methods in terms of Normalized Discounted Cumulative Gain (NDCG), at position 5 and position 25 (an estimate of first page length in a search session) on two product search datasets, both with more than 50,000 queries. One dataset is an open dataset made available during a benchmarking activity at CIKM  2016, the other dataset is a proprietary dataset obtained from a large European e-commerce platform ~(Sections~\ref{sec:background} and \ref{sec:experimental-setup}). 
	
Our main experimental finding (detailed in Section~\ref{sec:results}) is the following: modern learning to match methods are able to make an improvement of $134.46\%$ in terms of NDCG at position 5 of the list, on top of a lexical baseline that is based on BM25 on the CIKM 2016 dataset, while on the proprietary dataset this improvement is not bigger than $29.93\%$.
We attribute this finding to the fact that in our proprietary dataset almost all the items presented on the first result page have a high lexical overlap with the query, while in the public dataset the word overlap between query and product descriptions is about $1.8\%$, which is very low~\cite{wu2017ensemble}. This implies that for the public dataset there are more opportunities for semantic methods to prioritize some items over the others.

We find a high degree of correlation between the performance of the learning~to~match methods on our two datasets. Except for a special case, ARC-I, we see the same models corresponding to the top 5 scores achieved for both datasets. We find that models that have been specifically designed for short text matching, such as MV-LSTM and DRMMTKS, are consistently among the top three methods in all experiments, while the performance of a BERT-based model is mediocre. Moreover, we show model behavior regarding different aspects of queries, namely query length and popularity, and explain similarities and differences between the two datasets. We also look deeper into the queries for which either of the two matching methods, lexical or semantic, is preferred and discuss their characteristics. We found that for most queries in both datasets semantic matching can improve the ranking, however, since the fraction of queries hurt is substantial, we conclude that query-dependent selection of matching function would be beneficial.

\header{Implications for e-commerce practitioners} The effectiveness in terms of NDCG is not the only criterion in selecting a learning-to-match model for a real world use case. As \citet{trotman-2017-architecture} point out, efficiency (both at training and inference time) is a major consideration for product search on e-commerce platforms. We therefore analyze our results with a focus on choosing a suitable model for production deployments, and discuss how this choice is influenced by the trade-off between computation time and model performance~(Section~\ref{sec:impl}). We have seen that ARC-I provides a good balance between, on the one hand effectiveness improvements over and above a lexical baseline, with minimum effort required for fine-tuning, and on the other hand efficiency.  

\section{Related work}
\label{section:relatedwork}

\header{Product search}
Many approaches have been proposed for product search, ranging from adaptations of general web search models~\citep{duan2013supporting} to using versions of faceted search to speed up browsing of products~\cite{vandic2013facet,vandic2017dynamic}. 
To help select optimal approaches, \citet{sondhi2018taxonomy} propose a taxonomy for queries and costumer behavior in product search. 
Independent of the type of query, it is customary to consider a cascade of two or more steps in producing a \ac{SERP}: first we retrieve all potentially relevant items; then, using one or more re-ranking or learning to rank steps, we decide which items to put on top~\citep{trotman-2017-architecture}. 

There are specific challenges involved in applying learning to rank to product search; \citet{apps} study these in an e-commerce setting.
The signals used for matching queries and products in this learning to rank setup are diverse.
E.g., \citet{wu2017ensemble} combine three types of feature: 
\begin{enumerate*}
\item statistical features (e.g., total show count, click count, view count, purchase count of a product);
\item query-item features (e.g., content-based query-description matching); and, finally, 
\item session features about co-clicked items in sessions.  
\end{enumerate*}
\citet{ludewig-2019-learning} also consider a broad range of ranking features in a hotel ranking setting, with features ranging from descriptive statistics and latent features to product and location features.
In this paper we focus, specifically, on query-item features and contrast their effectiveness for product search.

\header{Learning to match}
\label{sec:rel-ltm}
Many learning to match methods based on deep neural networks have recently been introduced and used in a range of retrieval tasks; see, e.g., \citep{mitra2017neural,onal-neural-2018} for overviews. 
In product search, in addition to \citep{wu2017ensemble,ludewig-2019-learning}, \citet{bell-2018-title} develop an e-commerce-specific learning to match function based on query specific term weights and \citet{zhang2019neural} use interaction data between queries and products contained in a graph, along with text embeddings generated by a deep learning to match model to rank a list of products. 
\citet{magnani-2019-neural} devise (deep) learning to match models for product search, based on different types of text representation and loss functions.

\header{Comparing learning to match methods}
Systematic comparisons of (deep) learning to match methods are rare.
Exceptions include work by \citet{linjordet-2019-impact}, who examine the impact of dataset size on training learning to match models. 

\citet{guo2019deep} summarize the current status of neural ranking models, as well as their underlying assumptions, major design principles, and learning strategies. They survey the published results of some neural ranking models for ad-hoc retrieval and QA tasks, but mention that it is difficult to compare published results across different papers since even the smallest changes in experimental setup can lead to significant differences.

\citet{brenner-2018-end-to-end} contrast the use of learning to match models on web search vs.\ on product search and find a complex trade-off between effectiveness and efficiency.
\citet{ludewig-2019-performance} benchmark four neural approaches in the context of session-based recommendation against a nearest neighbors-based baseline and identify important lessons for reproducibility.
\citet{yang2019critically} apply five neural ranking models on the Robust04 collection to examine whether neural ranking models improve retrieval effectiveness in limited data scenarios.

What we add on top of the previous work listed above is a systematic study of the effectiveness and efficiency of learning to match methods for product search.
To facilitate reproducibility, we use the methods implemented in the MatchZoo library~\cite{guo2019matchzoo} for our study, like~\citep{linjordet-2019-impact}, as well as a BERT-based baseline.
We summarize those methods in Section~\ref{sec:background}, and in Sections~\ref{sec:experimental-setup} and~\ref{sec:results} we detail our experimental setup and outcomes.


\section{Learning To Match Methods}
\label{sec:background}

We summarize the learning to match methods that we evaluate in our experimental study.
\citet{guo2016deep} propose a categorization of learning to matching models as follows: \emph{representation-based models} aim to obtain a representation for the text in both queries and documents; \emph{interaction-based models} on the other hand, aim to capture the textual matching pattern between input texts.
We follow this organizing principle.

\subsection{Representation-Based Models}

\header{DSSM} The Deep Structured Semantic Model~\cite{huang2013learning} maps text strings to a common semantic space with a deep neural network (DNN) that converts high-dimensional text vectors to a dense representation. Its first layer applies a letter n-gram based word hashing as a linear transformation to reduce the dimensionality of feature vectors and to increase the model's robustness against out-of-vocabulary inputs. Its final layer computes the cosine similarity between the embedding vectors as a measure of their relevance. This model is trained on clickthrough data to maximize the conditional likelihood of a clicked document given the query. 

\header{CDSSM} The Convolutional Deep Structured Semantic Model~\cite{shen2014learning} extends DSSM and adopts multiple convolution layers to obtain semantic representations for  queries and documents.
Its first two layers transform the input to a representation, based on word and letter n-grams. Next, it extracts local and global (sentence-level) contextual features from a convolution layer followed by max-pooling, before computing the final matching score like DSSM.

\header{MV-LSTM} The MV-LSTM~\cite{guo2018interpretable} captures local information to determine the importance of keywords at different positions. It leverages a Bi-LSTM to generate positional sentence representations. Its Bi-LSTM generates two vectors that reflect the meaning of the whole sentence from two directions when based on the specified position. The final positional sentence representation is obtained by concatenating these vectors. 
MV-LSTM produces a matching score by aggregating interaction signals between different positional sentence representations to take into account contextualized local information in a sentence.

\header{ARC-I} The ARC-I~\cite{hu2014convolutional} network employs a convolutional approach for semantic similarity measurements. It leverages pre-trained word embeddings, and several layers of convolutions and max-pooling to generate separate dense representations for queries and documents. Finally, it applies an MLP to compare the resulting vectors via a non-linear similarity function.

\subsection{Interaction-Based Models}

\header{ARC-II} Models that defer the interaction between inputs until their individual representations ``mature,'' like ARC-I, run the risk of losing information that can be important for matching, because each representation is formed without knowledge of the others~\cite{hu2014convolutional}. 
ARC-II~\cite{hu2014convolutional} addresses this problem by using interactions between query and document, so that the network gets the opportunity to capture various matching patterns between the input texts from the start. It learns directly from interactions rather than from individual representations. A first convolution layer creates combinations of the inputs via sliding windows on both sentences, so that the remaining layers can extract matching features.

\header{DRMM} \citet{guo2016deep} mention three factors in relevance matching: exact matching signals, query term importance, and diverse matching requirements, and design the architecture of their deep matching model (DRMM) accordingly. DRMM first builds interactions between pairs of words from a query and a document, and subsequently creates a fixed-length matching histogram for each query term. Next, the model employs a feed forward network to produce a matching score, and calculates the final score by a weighted aggregation of the scores of the query terms. 

\header{DRMMTKS} This model is a variant of DRMM provided by the MatchZoo~\cite{guo2019matchzoo} library. It is meant for short-text matching, and replaces the matching histogram with a top-$k$ max pooling layer.

\header{MatchPyramid} This convolution-based architecture views text matching as image recognition~\cite{pang2016text}. The model first constructs a word-by-word matching matrix by computing pairwise word similarities.  This matrix is processed by several convolutional layers to capture the interaction patterns between words, phrases and sentences. In the first layer, a square kernel of size $k$ extracts a feature map from the matching matrix, which is aggregated by max-pooling to fix the feature size. Repetitions of these layers produce higher-level features of a pre-defined size as the final embedding.

\header{K-NRM} The kernel-based neural ranking model employs kernels to produce soft-match signals between words~\citep{xiong2017end}.  Given a query and document, it constructs a translation matrix from word pair similarities. These similarities are based on word embeddings that are learned jointly with the ranking model. In the next layer, kernels generate $K$ soft-TF ranking features by counting soft matches between pairs of words from queries and documents at multiple levels. The model combines these soft-TF signals and feeds them into a learning to rank layer to produce the final score.

\header{CONV-KNRM} This model~\cite{dai2018convolutional} is a variant of KNRM and applies a convolution to represent n-gram embeddings of queries and documents, from which it builds translation matrices between n-grams of different lengths in a unified embedding space. Its remaining architecture is identical to KNRM.

\subsection{Hybrid Models}

\header{DUET} DUET is a duet of two DNNs that combines the strengths of representation- and  interaction-based models~\citep{mitra2017learning}. DUET calculates the relevance between a query and a document using local and distributed representations and for this reason, has been classified as a hybrid model~\citep{guo2019deep}. By local representation we mean properties like the exact match position and proximity while distributed properties are synonyms, related terms and well-formedness of content.

The local sub-network applies a one-hot encoding to each term, from which it generates a binary matrix indicating each exact match between query and document terms. This interaction matrix is fed into a convolution layer, and the output is passed through two fully-connected layers, a dropout layer and another fully-connected layer, which generates the final output. The distributed sub-network takes a character n-graph based representation~\citep{huang2013learning} of each term in the query and document. 

For the distributed part, DUET first learns non-linear transformations to the character-based input by  applying a convolution layer on both queries and documents. This step is followed by a max-pooling step, whose output is the processed by a fully-connected layer. The matrix output for the documents is passed through another convolution layer. It then performs the matching by the element-wise product of the two embeddings. Next, it passes the resulting matrix from the previous step to fully-connected layers and a dropout layer until it obtains a single score. These two sub-networks are jointly trained as one deep neural network. 

\header{BERT} 
BERT is a deep bidirectional transformer architecture pre-trained on large quantities of text~\citep{devlin2018bert}, which has recently achieved state-of-the-art results on ad-hoc retrieval~\citep{macavaney2019cedr,qiao_understanding_2019}. BERT encodes two meta-tokens, namely [SEP] and [CLS], and uses text segment embeddings to simultaneously encode multiple text segments. [SEP] and [CLS] are used for token separation and making judgments about the text pairs, respectively. In the original pre-training task [CLS] is used for determining whether the two sentences are sequential, but it can be fine-tuned for other tasks. In our experiments we adopted BERT for classification, and to this end, a linear combination layer is added on top of the classifier [CLS] token~\citep{macavaney2019cedr}.


\section{Experimental Setup}
\label{sec:experimental-setup}
In this section, we introduce our research questions, and describe the two datasets we use for the experiments. We also describe the model tuning procedures and evaluation methodology.


\header{Research questions} Our experimental study aims to answer the following research questions:

\headerl{RQ1} \textit{How do learning to match models perform in ranking for product search compared to each other and to a lexical matching baseline?}  

\vspace{1mm}\noindent We consider the following aspects while investigating RQ1:

\begin{itemize}[nosep,leftmargin=*]
  \item \textit{How do learning to match models perform for different types of query in terms of length and popularity?}
  \item \textit{What is the per query score difference of the best semantic model compared to the lexical matching baseline?}
  \item \textit{How does BERT perform on short, unstructured text from product search descriptions?}
\end{itemize}

\noindent%
By answering these questions we want to provide insights into the usefulness of these models in a comparable setting for product search. In product search it is important to know the model behavior for different types of query. For example, some e-commerce platforms may prefer to respond as effectively as possible to popular queries even if it yields low performance for long-tailed ones. On the other hand, some might prefer a model that is quite robust to different aspects of queries like length and popularity. In the last question we want to investigate the impact of BERT pre-trained weights on the data we collected from a product search engine, given the fact that the documents used in pre-training BERT are well-structured sentences, but in our case, the majority of documents (product descriptions and queries) can be considered as phrases or combinations of keywords.

Based on the experimental results, we aim to assist data scientists in e-commerce scenarios by focusing on two additional research questions, which concern the choice of a suitable model for e-commerce scenarios:

\headerl{RQ2} \textit{Can we come to a general conclusion about which category of models, interaction-based or representation-based, to prefer for the product search task?}  

\vspace{1mm}\noindent%
Most e-commerce companies have a high number of searches per second, therefore, it is important for them to be able to conduct some part of the model training offline for efficiency purposes. Representation-based models can generate embeddings for documents separately from queries (offline), which is an important advantage over interaction-based models in an online setting. This motivates a comparison of these model classes to obtain an understanding of their (dis)advantages for e-commerce practitioners. 

Furthermore, every production deployment of machine learning has to take the cost incurred by model training and inference into account, which often has to be traded off against the business benefits provided by the model~\cite{liberty2020elastic}. The time for inference is also crucial, as the response latency of a model has a major impact on its online performance~\cite{arapakis2014impact}. We therefore ask the following research question:

\headerl{RQ3} \textit{Will the choice of preferred models change when we take training time, required computational resources and query characteristics into account?}  

\vspace{1mm}\noindent%
By targeting this question, we aim to assist e-commerce practitioners with the decision which models to select as candidates for their production use cases.

\header{Datasets} We conduct experiments on two datasets, of which one is publicly available and the other one is proprietary. The basic statistics of the two datasets are shown in Table~\ref{data_table}. 

\headerl{CIKM~2016} Our first dataset is the publicly released dataset from track two of the CIKM Cup 2016.\footnote{\url{https://competitions.codalab.org/competitions/11161}}
This dataset contains six months of anonymized users' search logs including query and product description tokens, clicks, views and purchase records on an e-commerce search engine from January 1st, 2016 to June 1st, 2016. The dataset contains additional product metadata such as product categories, description and price. In the original split of the data, the last query of each session is marked as test sample, so we do not have user interaction signals for these queries. In addition to search sessions that come with queries, this dataset also contains browsing logs that are query-less. We ignore this part of the data in our experiments, since we are interested in query/document matching based on text, and we study its impact in a SERP re-ranking task. As a consequence, since the results achieved here are only based on text matching on query-full queries, they are not comparable to studies in which the whole dataset is used to improve the ranking such as~\citep{zhang2019neural,wu2017ensemble}.

\begin{table}[t]
    \centering
    \caption{Basic statistics of our datasets.}
   \begin{tabular}{lrr}
     \toprule
        & \bf CIKM 2016  & \bf Proprietary \\
      \midrule
      \#queries &  51,888 &  53,474 \\

      \#unique queries	& 26,137&40,125 \\

      \#unique presented products & 37,964 &  214,778 \\
      \#clicks &36,814&63,859\\
      \bottomrule
   \end{tabular}
   \label{data_table}
\end{table}

\headerl{Proprietary dataset} Our second dataset is extracted from the search logs of a large, popular European e-commerce platform.
The main language of the dataset is Dutch.
This dataset contains sampled queries from ten days of users' search logs. 
We leverage the first nine days as training data and the last day queries as test data. 
For each query, we know the items rendered on the first result page.
We only use the title for each item, which contains a short description of the product to be consistent with the CIKM~2016 data.
We label positive matches between queries and items according to observed click-through data, and remove sessions without clicks from the dataset~\cite{joachims2002optimizing}. In the preprocessing step, we remove punctuation marks, HTML tags, and other unknown characters from the text. Also, we lowercase all the tokens.

\header{Model tuning} We experiment with models from the well established MatchZoo~\cite{guo2019matchzoo} library,\footnote{\url{https://github.com/NTMC-Community/MatchZoo}} which itself is based on Keras and TensorFlow.  The MatchZoo library contains a tuning module that fine-tunes the models based on pre-defined model-specific hyper-parameter spaces~\citep{guo2019matchzoo}. In the tuning process parameters are sampled from the hyper-space, but this sampling is not random, the scores of past samples will have an effect on the future selection process in a way that it yields a better score. The tuner uses Tree of Parzen Estimators (TPE)~\cite{bergstra2011algorithms} to search the hyper-space. We start tuning from the default parameter values provided by MatchZoo, and select hyper-parameters for all models based on a fixed validation set in $50$ rounds. 

For the experiments with BERT, we used the implementation from Contextualized Embeddings for Document Ranking (CEDR)\footnote{\url{https://github.com/Georgetown-IR-Lab/cedr}} provided by~\citet{macavaney2019cedr}. We employed BERT base model (12 layers) with multilingual weights as well as a (Dutch) monolingual version called BERTje~\citep{de2019bertje}, which is trained on a large and diverse dataset of 2.4 billion tokens. We conducted separate experiments with both sets of weights in this work. We also added gradient clipping and warm-up steps from the HuggingFace transformer~\citep{Wolf2019HuggingFacesTS} implementation to improve the performance.\footnote{\url{https://github.com/huggingface/transformers}}

\header{Evaluation setup} The cascade model~\cite{craswell2008experimental} assumes that users scan items presented in a \ac{SERP} one by one, from the top of the list, and the scan will continue after observing non-relevant items but stops after a relevant item is found. 
Motivated by this assumption, we only consider the items above the last clicked product in the list as well as two items below that. This approach additionally helps to reduce the data size while balancing the number of positive and negative samples in our data.
This principle is applied to both datasets, and we use the same maximum number of epochs with early stopping for training all the models. Moreover, since we labeled our proprietary data only based on clickthrough information, we treat clicks and purchases identically for the CIKM dataset. 
While we consider the items to be presented in a list, it is common for e-commerce websites to use a grid view to display the products. In this case, users' examination behavior can be different from the cascade model we use in this study.

\header{Evaluation metrics} We report NDCG at two cut-offs: 5 and 25. We decided for a cut-off of 5 because the top items returned for a query are important to capture a user's attention; we choose 25 which is the maximum number of results per query in both datasets, and it is a good estimate of the items shown on the first page of an e-commerce website.


\section{Performance of Learning To Match Methods for Product Search}
\label{sec:results}

In this section, we seek to answer the research questions mentioned in Section~\ref{sec:experimental-setup}. For this purpose we study the performance of different learning to match models in a comparable setting. 

We first address RQ1: \textit{How do learning to match models perform in ranking for product search compared to each other and to a lexical matching baseline?} by determining the best-performing method or group of methods in bridging the vocabulary gap for product search. 

The overall performance of the learning to match methods on our datasets is summarized in Table~\ref{model_results}. Here, we re-rank an original ranked list obtained by a lexical matching method as first step in a two-step retrieval cascade. For the CIKM data, the original ranking comes from BM25, so it is only based on matching between query, and product title. For the proprietary data, we omit signals involved in the production ranking which are not related to lexical match, so that we obtain a similar baseline as the one we have for the public data. As a result, the ranking produced by the lexical baseline is purely based on matches of the query and the title, which enables us to compare the results to the BM25 baseline provided for the public dataset.

\begin{table}[t]
    \caption{Performance of MatchZoo models on both datasets in terms of NDCG at position $5$ and $25$.}
  \begin{tabular}{l SS SS}
    \toprule
    \multirow{2}{*}{Model} &
      \multicolumn{2}{c}{CIKM data} &
      \multicolumn{2}{c}{Proprietary data} \\
      \cmidrule(r){2-3}
      \cmidrule(r){4-5}
      {}  & {NDCG@5} & {@25}  & {@5} & {@25}  \\
      \midrule
    Lexical  & 0.148 &0.343& 0.314 &   0.474\\
    \midrule
    MatchPyramid  & 0.152 & 0.347& 0.287 & 0.454 \\
    CDSSM  & 0.314& 0.452 & N/A & N/A \\ 
    ARC-II  & 0.320 & 0.458& 0.334 & 0.488 \\
    ARC-I  & 0.326 & 0.462 & \bfseries 0.408 & \bfseries 0.549 \\
    DRMM & 0.331 & 0.464& 0.288 & 0.455 \\
    DSSM & 0.334 & 0.467  & N/A & N/A  \\
    KNRM  & 0.341 & 0.472& 0.337 & 0.490 \\
    DUET  & 0.345 & 0.473& 0.350 &  0.500 \\ 
    MV-LSTM & 0.342 & 0.474 & \bfseries 0.408 & \bfseries 0.549\\
    CONV-KNRM  & 0.347&0.476 & 0.349 & 0.498 \\ 
    DRMMTKS &\bfseries 0.347 & \bfseries 0.477 & 0.345 & 0.498\\
    ‌Best-BERT & N/A & N/A &0.340 & 0.493 \\
    \bottomrule
  \end{tabular}
  \label{model_results}
\end{table}

\header{Results for the CIKM2016 dataset} For the CIKM dataset, all learning to match models outperform the lexical match baseline. 
The score achieved by MatchPyramid is almost the same as the baseline, but other models perform $112.16\%$ to $134.46\%$ better than BM25 in terms of NDCG@5 for the public dataset. It is worth mentioning that, although, the differences in scores might not be noticeable, they indicate improvement for many queries. In other words, the $0.001$ difference between the scores achieved by CONV-KNRM and DRMMTKS at position 25, means better NDCG for $3.22\%$ of test queries. DRMMTKS performs better than the baseline for $36.02\%$ of test queries.
This confirms that semantic matching can indeed improve the matching of query/item pairs in product search as well as more general ranking tasks, even in the absence of well-structured sentences or long documents.

\header{Results for the proprietary dataset} Not all semantic matching models outperform the lexical matching baseline for the proprietary dataset. Specifically, MatchPyramid and DRRM achieve lower results for this dataset. On average, the spread of the results we get for this data is smaller than for the CIKM dataset. In the latter the improvement made by the best performing model -- DRMMTKS -- is roughly $134.46\%$ better in terms of NDCG@5 compared to the lexical baseline, which implies that in this case, semantic matching can greatly help ranking most relevant items on top of the list. However, the corresponding improvement of our learning to match methods for the proprietary dataset is not bigger than $29.93\%$ when we take ARC-I/MV-LSTM as the best performing semantic methods. If we compare MatchPyramid's score as the lowest, to ARC-I/MV-LSTM  the gain is $42.16\%$ which is still way smaller than what we see in the public data. It should be noted that, although we report the same score for ARC-I and MV-LSTM, they perform differently for $0.4\%$ of test samples. Since the difference is marginal, it is not visible in 3 decimal digits.

We attribute the lower impact of the semantic matching methods on the proprietary dataset to the fact that almost all the items presented on the first page have a high lexical overlap with the query. In other words, the diversity of the first page, if we only consider contextual aspects of products (titles) as the source of diversity, is much smaller compared to the public dataset. This implies that for the public dataset there are more opportunities for semantic methods to prioritize some items over others. Besides, the chronological split of our proprietary dataset makes it a more challenging case than the public dataset.

In general, we observe that models like MV-LSTM and DRMMTKS are consistently among the top performing methods in all experiments, which we attribute to the fact that these models have been specifically designed for short text matching. The average length of queries in our public dataset is $3.1$ and the average length for product descriptions is $4.8$, which both are very short.

Note that we could not successfully finish a run of CDSSM and DSSM on the proprietary dataset, due to out-of-memory issues with the respective MatchZoo implementations. We plan to address this  issue in future work.
 
One aspect of RQ1 is to investigate the performance of BERT-based models on short, unstructured text from product search logs. As indicated in Table~\ref{model_results}, our best performing BERT-based model (Best-BERT) which employed ``bert-base-multilingual-uncased" pre-trained weights, and early stopped after 10 patience steps, is not among the top ranked models for our proprietary dataset. In Section~\ref{sec:experimental-setup} we mentioned that we also considered another version of BERT named Bertje, however, since we have English terms mixed in with the (predominantly) non-English text in product titles and queries, multilingual weights performed better than a monolingual model. The NDCG@25 achieved with Bertje pre-trained weights on our dataset is $0.488$ while the score achieved from multilingual weights, in the same setting, is $0.493$. It is worth mentioning that to study the impact of fine-tuning on our dataset, we once applied the model on the test data without any fine-tuning. The performance we obtained is $0.445$ which implies the effectiveness of fine-tuning.  

We conjecture that one of the main reasons behind this poor performance from state-of-the-art BERT is the fact that the text it is pre-trained on is very different from the text we have in product search; more investigations are needed to support this conjecture.

\subsection{Performance for Different Types of~Queries}
\label{sec:results:lenght-popularity}

Next, we drill down into the experimental results to investigate an additional aspect of RQ1: \textit{How do learning to match models perform for different types of query in terms of length and popularity?}

\begin{figure}
\centering
  \includegraphics[width=\linewidth]{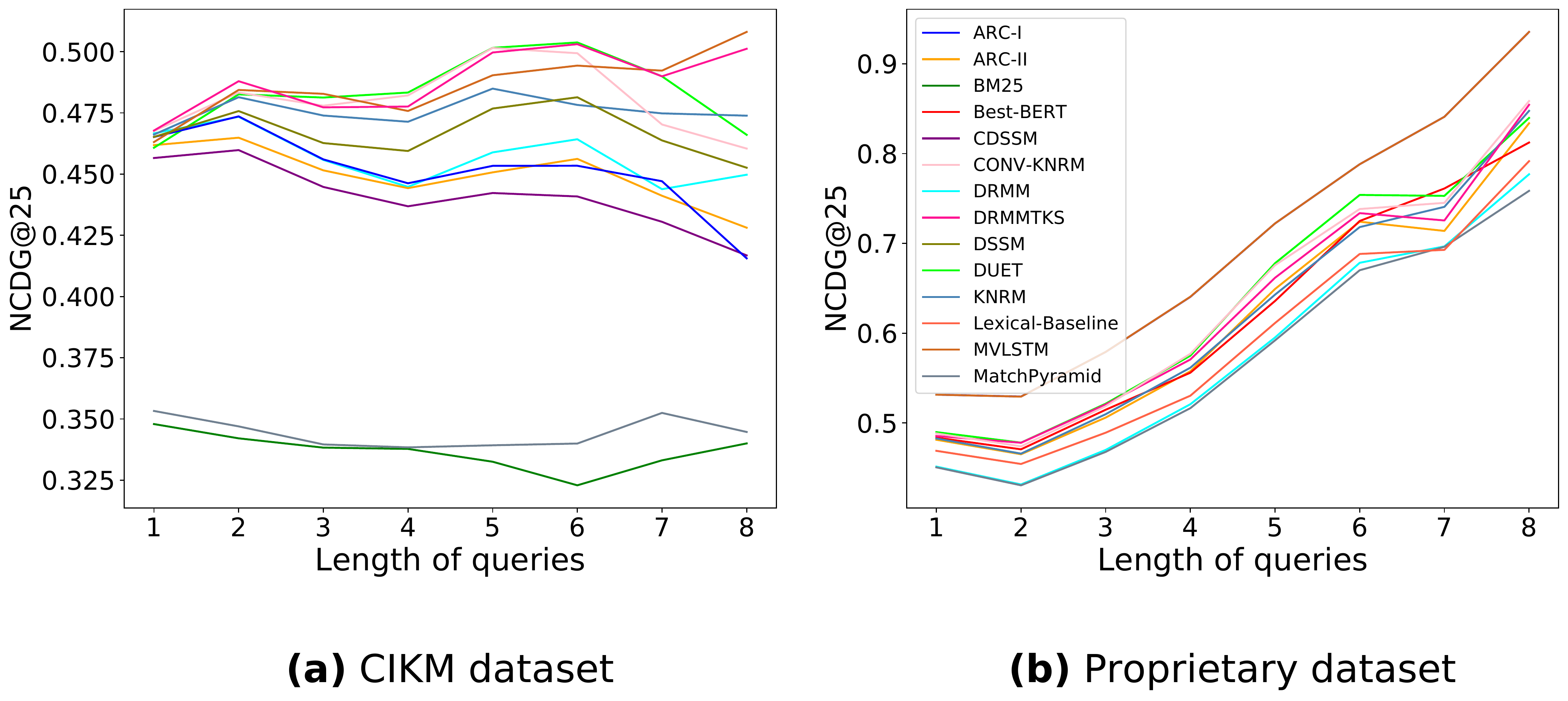}
  \caption{Ranking performance for varying query length.  On the X-axis we see the length of the query, and Y-axis indicated the average NDCG at position 25 per queries of a specified length.}
  \label{fig:query_len}
\end{figure}

\header{Query length} Figure~\ref{fig:query_len} depicts the ranking performance of all models under varying query lengths. Most of the queries in our datasets contain a single word only, but there are a few very long queries with more than 75 words in the CIKM data and smaller queries of 17 words in our proprietary data. Note that we restrict ourselves to query lengths up to 8 words for which we have a sufficient number of samples in our datasets.

\begin{figure}
\centering
  \includegraphics[width=\linewidth]{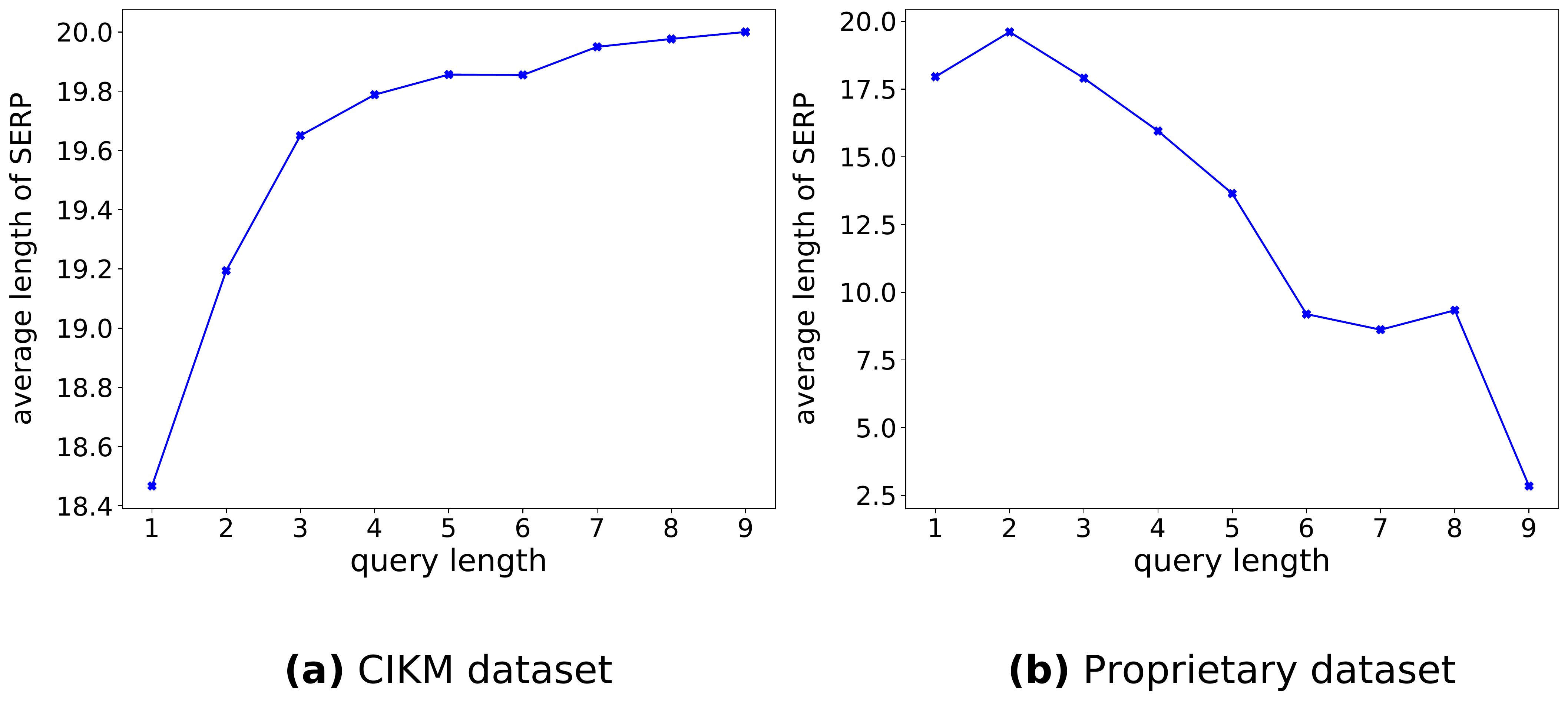}
\caption{Average number of items presented on a SERP for different lengths of queries. 
}
  \label{fig:serp_len}
\end{figure}

For the proprietary dataset (Figure~\ref{fig:query_len}b), we observe that as the query length increases the matching performance increases too. All the models follow the same trend. Figure~\ref{fig:serp_len} shows the average number of items presented in response to queries in both datasets. In general this number is larger for the CIKM data and we can see that the length of SERP increases by the length of queries. In our proprietary dataset, however, longer queries result in fewer items, which is attributed to the fact that in most cases these long queries are the exact descriptions of specific products which are already known by the users. Since in these cases, the search engine can precisely retrieve the intended products, the users can be easily satisfied, which is visible in high scores of NDCG for these queries. Unfortunately, we do not have access to the actual content of the CIKM queries, so we cannot further interpret the behavior of this data.   

For the  CIKM dataset (Figure~\ref{fig:query_len}a) however, we cannot arrive at a reasonable conclusion regarding the relationship between query length and performance. Since the terms in this dataset are hashed, it is difficult for us to investigate the performance of different methods based on query length as we do not know the original terms. When comparing different models, KNRM seems more robust to query length compared to the other models, and CONV-KNRM also performs quite consistently for shorter queries. 

\begin{figure}
\centering
  \includegraphics[width=\linewidth]{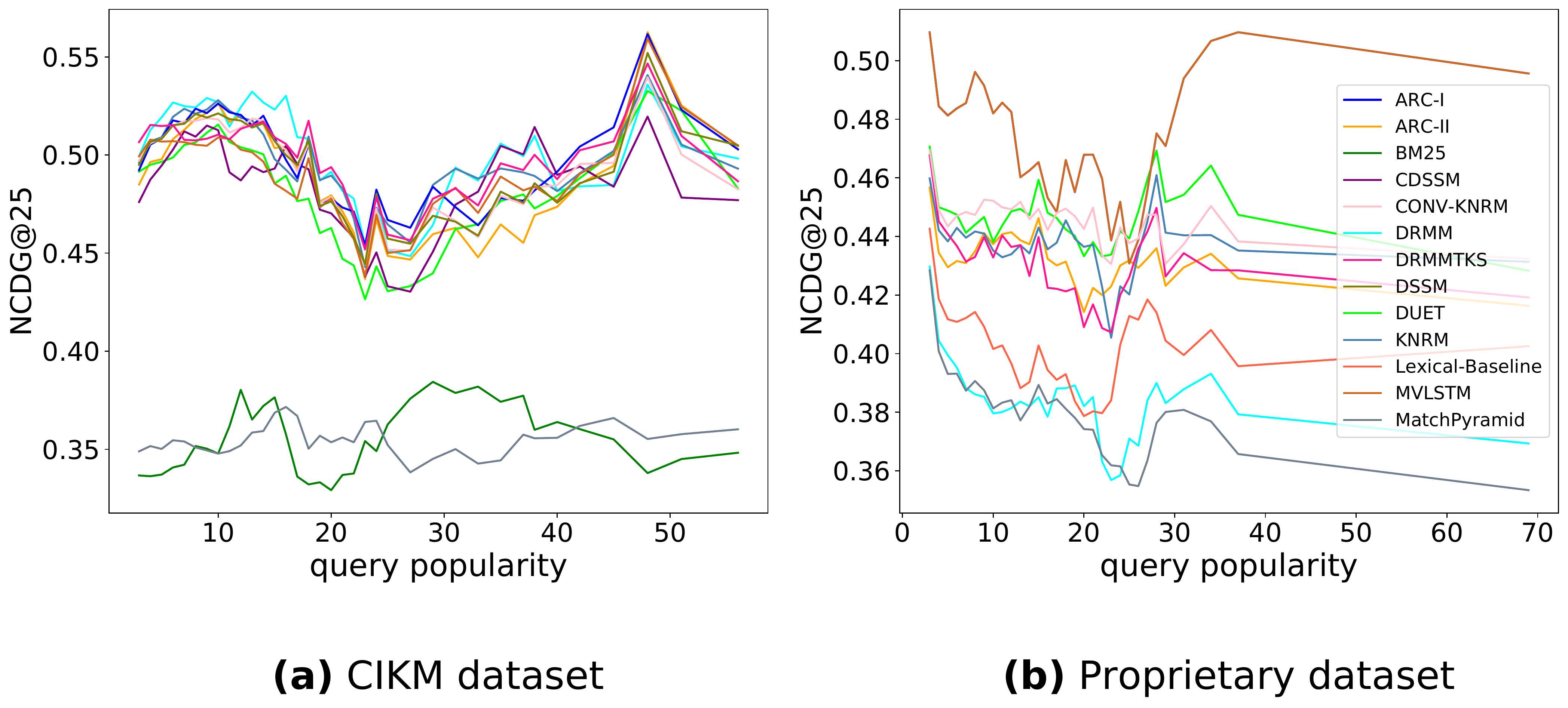}
 \caption{The behavior of models based on query popularity. The flow is quite the same in all cases and all of the models tend to perform better for more frequently seen queries.}
  \label{fig:query_pop}

\end{figure}

\header{Query popularity} Figure~\ref{fig:query_pop} depicts the results based on the popularity of queries, i.e., the number of times a query is repeated throughout our dataset. We again only include popularity values for which we have a sufficient number of samples. The X-axis indicates the popularity of the queries, and the Y-axis denotes the average NDCG at position $25$. 

Interestingly, we observe a ``valley'' in the middle for both datasets. We can explain what we see in Figure~\ref{fig:query_pop}b in three steps: starting from leftmost part of the plot, it contains less popular queries which are usually longer than the popular ones, and from what was indicated in Figure~\ref{fig:query_len}b, we know that it is easier for the models to rank items for these types of queries. That is why we see a relatively high performance at this part. However, as the popularity increases the queries get shorter. The middle part contains queries that are repeated between 10 to 15 times, we encounter some shorter queries, which are more challenging for the models, and are not repeated often enough for the models to pick up the patterns between the pairs of these queries with the large number of associated items. This situation gets more difficult since based on the nature of the original production ranking function which was employed during logging of our proprietary data, we often see that the retrieved list for a query can vary a lot from day to day. As we move further in Figure~\ref{fig:query_pop}b toward the most popular queries which we consider to be short, the performance improves.  The reason for this observation can be that popular queries get repeated over and over again, the lists of items presented for them converge, and the models can learn the relationship between the pairs. As expected, we see that the models generally perform better for more popular queries.

\subsection {Per Query Score Difference between the Best Semantic Model
and the Lexical Matching Baseline}

Next, we focus on the final aspect of RQ1: \textit{What is the per query score difference of the best semantic model compared to the lexical matching baseline?}

\begin{figure}
\centering
  \includegraphics[width=\linewidth]{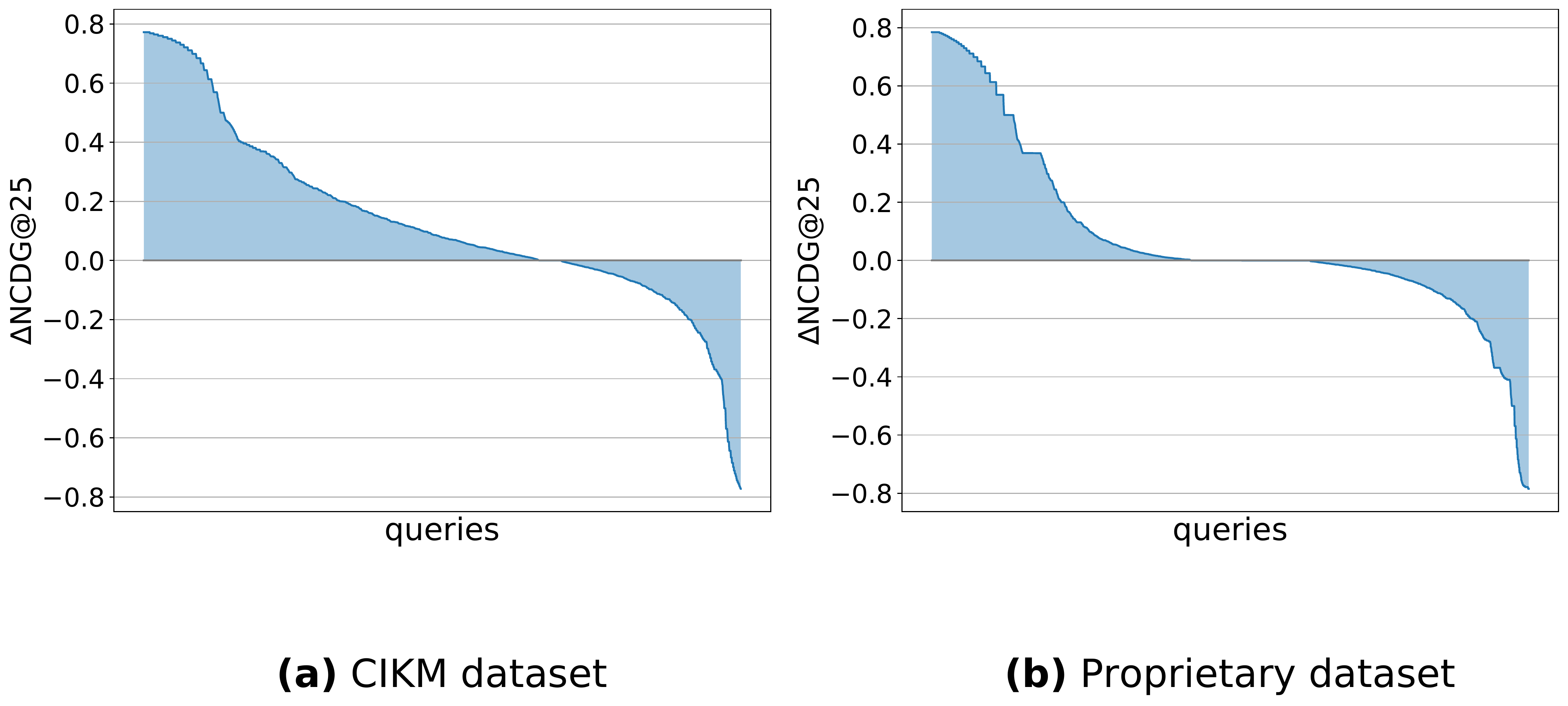}
  \caption{Per-query paired differences between the best semantic model and lexical baseline for models trained on each dataset and evaluated on the test sets. The Y-axis indicates $\Delta$NDCG at position 25 of ranking between best semantic model and lexical baseline
 The X-axis lists the queries in the referenced dataset in decreasing order of $\Delta$NDCG such that queries for which semantic model performs better are on the left and vice versa for the lexical model on the right.}
 \label{fig:semantic_lexical_dist}
\end{figure}

Figure~\ref{fig:semantic_lexical_dist} shows the difference between the best performing semantic model and the lexical match per query for each of the datasets. The best performing semantic model for the CIKM dataset was DRMMTKS, while ARC-I and MV-LSTM performed best for the proprietary data. The Y-axis shows the difference in NDCG at position 25 of the best performing model to the lexical baseline for all test queries. The X-axis lists the queries in decreasing order of $\Delta$NDCG such that the queries for which the semantic model performs better are on the left and vice versa for the lexical model on the right. 
Queries that benefit from semantic matching have a positive value on the Y-axis while those that prefer lexical matching have a negative value.
The plots indicate that semantic matching improves the ranking for most of the queries.
This is more obvious in Figure~\ref{fig:semantic_lexical_dist}a, considering that semantic matching has more influence on the CIKM data than for the proprietary data, which is also presented in Table~\ref{model_results}. Although in Figure~\ref{fig:semantic_lexical_dist}b this difference is not as visible, the area under the curve for the upper part is $~1.2$ times bigger than the part below the X-axis. This suggests that query-dependent selection of matching function would be beneficial.

In the case of the CIKM data, it is hard to interpret the queries from the uppermost and bottom points of the plot, since we do not have access to the actual content of the queries. However, for the proprietary data, we can analyze these respective queries. We observed that, analogous to the public dataset, there is no meaningful difference in terms of the length and the popularity of these queries, but the words in the queries for which semantic matching performs best are more general and commonly used than the words, which we see in the other group. For example, among the queries for which semantic method vastly outperforms lexical matching, we encounter queries like ``\emph{wireless earbuds}'', and  ``\emph{lego}'', which are closer to a category name than to one specific product. On the other hand, we have examples like ``\emph{anneke kaai},''  ``\emph{buzzed}'' and ``\emph{Stephan Vanfleteren}'' for queries with a higher NDCG achieved by the lexical matching baseline, these are mostly proper nouns or specific items. 

There are some queries that are repeated in both groups. In other words, we can see sessions with the same query, while in one session lexical matching performs better, in the other one semantic matching provides a better ranking. This is because of different relevance judgments from different users based on personal preferences. Examples of these queries are ``\emph{star wars lego}'' or ``\emph{perfume}'' of different brands. In these cases, all the products rendered in SERP contain the exact words from query, so the only factor that makes a user click on an items is personal preferences and possibly the position of the product in the list. When these preferences vary from user to user there is no discriminating signal that the semantic models can capture to prioritize one item over the other for future queries. 

When looking at per query best/worst performances of the semantic matching model and the lexical baseline on our proprietary dataset, we see that for $62.3\%$ of the queries both models either perform accurately or poorly. However, in $15.9\%$ of the cases, we have queries for which the performance of semantic matching is high, while lexical matching does not perform accurately. On the other hand, we see that the opposite behavior, i.e., where the lexical matching baseline outperforms semantic matching, is much rarer ($7.9\%$). 

Again it is interesting to look at some examples: among the queries for which semantic method is preferred, we can see queries expressed in general terms like ``\emph{woonkamer klok},'' ``\emph{tractor hout}'' or ``\emph{panty met print},'' which means ``living room clock," ``tractor wood," and ``panty with print," respectively. These queries do not match to any specific item, as they are more exploratory queries. Among the queries that do not benefit from a semantic method and prefer a lexical match, we again mostly see queries with proper nouns, and more interestingly combinations of numbers like \emph{11 400 700} which matches the sizes of some charging cables. For these queries, the user exactly knows what he/she is looking for in the product catalog and we see that these terms match to parts of titles.


\subsection{Further Considerations}
\header{Impact of word embeddings} For some of the models, the word embeddings used for initialization are more critical than for others. For example, DRMM creates a similarity matrix based on the word embeddings at the beginning, and does not update the embeddings in the training process (like MV-LSTM does, for example). In our experiments, we have all models start with a random initialization in order to have an identical setting for both datasets to make the results comparable. However, as a result we encountered very poor performance for DRMM as one of the worst performing models with NDCG scores of $0.331$ and $0.288$ at position $5$ for the public and proprietary data, respectively, which contrasts other studies, where it proved to be one of the strongest models for different tasks~\cite{yang2019critically, guo2019deep}. 

We ran an additional experiment for DRMM, ARC-I and ARC-II in order to investigate the impact of the leveraged word embeddings on their performance. For training these models we experimented with both a Word2Vec model learned from a large corpus of general text in the same language as our proprietary data, and a Word2Vec model learned from the text of a corresponding proprietary product catalog and our training queries. However, we did not observe meaningful improvements over using pre-trained embeddings. 

The problem with using embeddings learned from general text is that queries and product descriptions of our proprietary data contain both English and non-English text, so when using only non-English embeddings the model misses plenty of words. To solve this problem we also trained a Word2Vec model on the product catalog and training queries. However, we found it difficult to balance the amount of product descriptions and queries to achieve robust weights from the Word2Vec model.


\section{Implications for E-Commerce}
\label{sec:impl}

Experiments into position bias in e-commerce settings have shown that customers are prejudiced towards the first few results~\citep{hofmann-2014-effects}. It is customary to rank products primarily based on the popularity of a product without taking semantics into account~\citep{trotman-2017-architecture}. However, user studies and analyses of interaction logs reveal that customers use various queries with subtle differences to search for the same product set that should lead to different rankings~\citep{sondhi2018taxonomy}. Adding semantics, to query understanding and to product ranking, should result in a better ranking. From the point of view of generating semantically more meaningful ranked lists of products than purely ranking by popularity, our experiments in Section~\ref{sec:results} suggest that we should consider models like ARC-I, MV-LSTM and DRMMTKS. 

But effectiveness as measured in terms of NDCG is not the only criterion in selecting a learning to match model for a real world use case. As \citet{trotman-2017-architecture} point out, for  product search on e-commerce platforms, efficiency is a major consideration: both efficiency at training time and at inference time. In order to accommodate for the special considerations in production use cases, we analyze our results in this section. This discussion can be leveraged as a starting point for deciding which model to choose for a real world deployment by e-commerce practitioners. We focus on the question of which model family to choose~(Section~\ref{sec:choice}) and how this choice is influenced by the trade-off between computation time and model performance~(Section~\ref{sec:cost}).

\subsection{Choosing between Representation-Based Models and Interaction-Based Models}
\label{sec:choice}

We now discuss RQ2: \textit{Can we come to a general conclusion about which category of models, interaction-based or representation-based, to prefer for the product search task?}

The motivation for this question is as follows. Most e-commerce companies have a high number of searches per second, and are at the same time continuously expanding their total number of products, partners and customers. This results in a lot of new product and interaction data per day, as well as an ever shifting catalog of products. As a consequence, we require a model that can be extended with new data and is able to rank products that have not been seen with a particular query before. 

Representation-based models can generate embeddings for documents separately from queries and we can cache these embeddings for efficiency purposes. In these models the embeddings for query and product are not dependent on each other unlike interaction-based models where query and product are linked. This allows us to easily compute the embeddings for all products and popular queries offline. For interaction-based models, even if we manage to cache the representations of top items retrieved for popular queries, in the case of new products or changed items (possible changes in product description or other contents that we might use), we again need to compute online which can be very time consuming. 

Thus, we have a preference for representation based models. Based on the results in Table~\ref{model_results} interaction-based models, namely DRMMTKS and CONV-KNRM are the two best performing models for the public dataset with NDCG@25 of $0.477$ and $0.476$ respectively. However, the representation-based model MV-LSTM is in the third rank with the score of $0.474$, which is a marginal decrease compared to the two interactive models. Furthermore, for our proprietary dataset the best performance is from representation-based models, ARC-I and MV-LSTM, which is fortunate for practitioners. In summary, we would recommend representation-based models for production deployments in general, due to the discussed operational advantage of being able to incorporate new query and production representations easily~\citep{guo2019deep}.

\subsection{Training Cost and Inference Speed}
\label{sec:cost}

Next, we discuss efficiency specific aspects of RQ3: \textit{Will the choice of preferred models change when we take training time, required computational resources and query characteristics into account?}
In the real world, resources are always limited; by answering this question we aim to assist e-commerce practitioners with the decision which models to select as candidates for their production use cases. Inference time is also important since the response time of the system has a huge impact on the user experience. 

We compare the training and inference time for the learning to match models on our proprietary dataset. We only provide this information for our proprietary data, since it contains raw search logs extracted from an e-commerce platform, and the chronological split of test/train samples is a more acceptable representation of the real world use cases.

\begin{figure}
\centering
  \includegraphics[width=\linewidth]{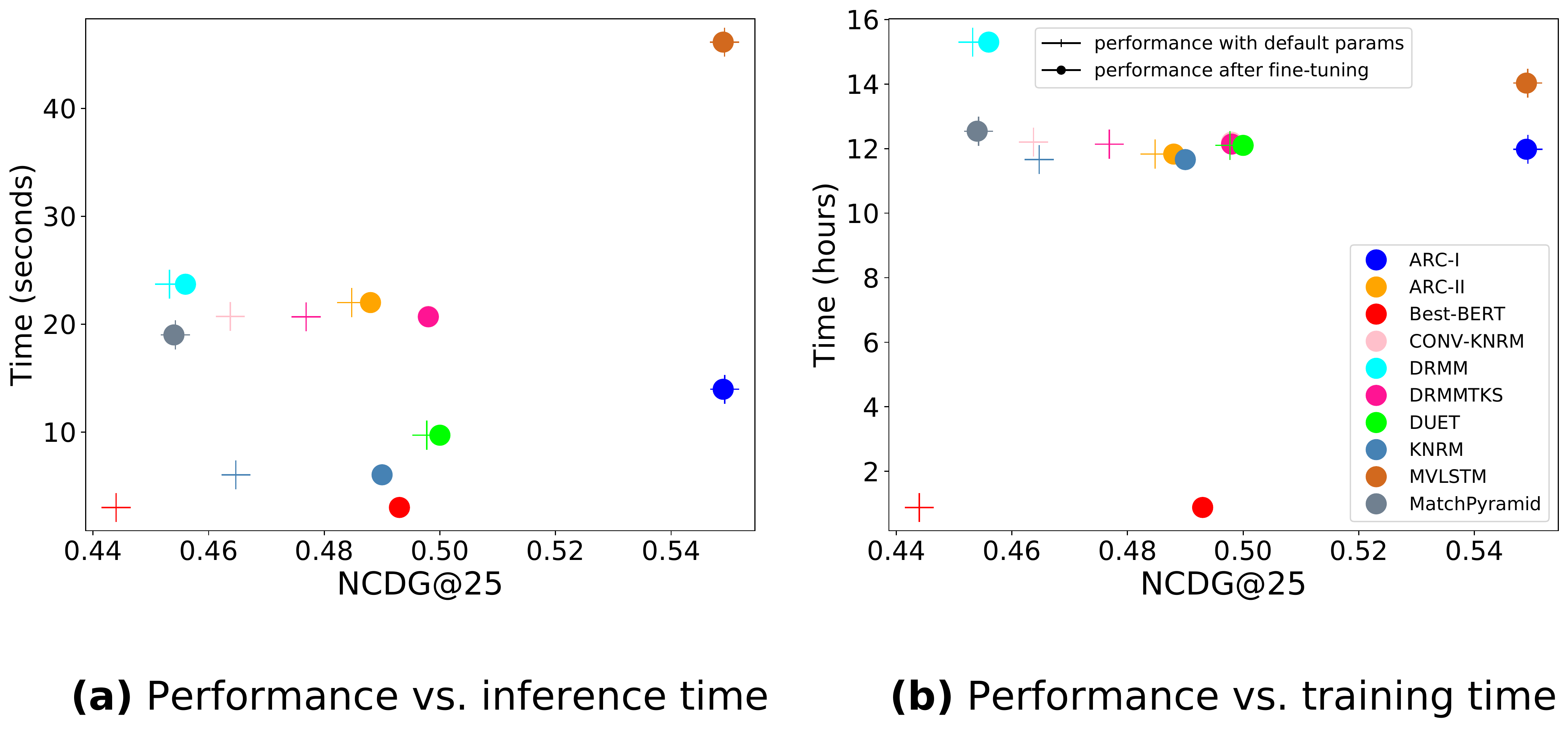}
  \caption{Ranking performance in comparison to training and inference time for the proprietary dataset. Both ranking performances achieved from the default hyper-parameters and the fine-tuned ones are depicted in this figure.}
  \label{fig:time_efficiency}
\end{figure}

Given that we have to trade-off computation cost and ranking performance, based on Figure~\ref{fig:time_efficiency}, we conclude that ARC-I and Best-Bert are the strongest candidates, since they have strong ranking performance (see Table~\ref{model_results}) while having low training and inference times. Considering ARC-I, it is really important that we could achieve the best performance on our dataset using this model with the default values for its hyper-parameters (Figure~\ref{fig:time_efficiency}). This means that compared to the other models, it is possible to obtain a sufficient performance using ARCI-I without the need to fine-tune the model.

\header{Memory consumption} In Section~\ref{sec:results} we mentioned that we could not successfully finish a run of CDSSM and DSSM on the proprietary dataset, due to out-of-memory issues with the respective MatchZoo implementations.
MatchZoo has specific preprocessing modules for these two models that include word hashing. This preprocessing step consumes a huge amount of memory, which makes it inapplicable of being applied on our proprietary data. 
 Although MatchZoo provides us with the option of not using word hashing for the preprocessing step of the training process, for the evaluation part it does not support the same setting. 
 From the experiments we observed that CDSSM was considerably slower than DSSM, but none of them can be considered proper choice for a big dataset like ours. 

It should also be noted that, although the MatchZoo implementation supports GPU computation, we observed a very low GPU usage for all of our models during training and evaluation time. In terms of average computational time spent on GPU non of the models exceeds $5\%$ GPU utilization during the whole process on any of our two datasets. Since we checked the functionality of the implementation with a QA dataset consisting of long documents, we can attribute this observation to the fact that our texts are too short to engage the GPU properly. We also contacted the MatchZoo team and they suggested to increase the batch size to a very big number to solve the issue, but since it could cause a lower ranking performance we did not follow this advice.

\subsection{Impact of Query Characteristics}

Finally, we discuss aspects of RQ3 with respect to query characteristics. Query characteristics are important for an e-commerce platform because it is important to know which model is preferred in which case. 
Length of the query and query popularity are two characteristics that differ per language and per device on the platform. 
The results from Section~\ref{sec:results:lenght-popularity} show that query length does have an influence on the model and it could be worth investing into various models for various settings if the average query length differs per setting. 
For example, from the analysis conducted, we see that customers who access the e-commerce platform through an app use shorter queries than customers who use a browser. 
E-commerce platforms could develop different models for different settings, although this might affect consistency to an unacceptable level. 

Query popularity is the second important characteristic.
Ideally, each query should result in the best ranking for the customer. 
However, popular queries are searched more often and thus these influence the revenue more than less popular queries.
A model should thus work well enough on the less popular queries and excellent on the popular queries with a certain trade-off, so MV-LSTM seems to be a good choice.
Having said that, there exist cases where less popular queries can equally influence the revenue (e.g., when they are issued after a popular query in a session). This should also be taken into consideration when selecting rankers for deployment.

\subsection{Recommendations for Deployment}

In terms of overall performance, we have seen that ARC-I provides a good balance between, on the one hand effectiveness improvements over and above a lexical baseline and on the other hand efficiency. Plus, the fact that it can perform well with the default configuration is another positive point.   

Next, we found that in terms of query length, most learning to match methods perform consistently across different query lengths on the CIKM dataset, with the results for MV-LSTM going up as query length increases; on the proprietary dataset, the performance of all methods consistently increases with query length. In terms of query popularity, we see consistent performance across different levels of popularity for all learning to match models on the CIKM dataset, but on the proprietary data we seen that some learning to match methods clearly benefit from increased popularity, including DUET and MV-LSTM.

Furthermore, in a side-by-side comparison between top performing learning to match methods and a lexical method, we see that substantial fractions of queries are helped by the learning to match method than are hurt, on both the CIKM dataset and the proprietary dataset, while the fraction of queries hurt is substantial. 
The latter suggest that query-dependent selection of a matching function would be beneficial.

Finally, we found that representation-based models provided the best trade-off between accuracy and efficiency.


\section{Conclusion \& Future Work}

In this paper, we have set up a comprehensive comparison of supervised learning to match methods for product search. We have considered 12 learning to match methods, considering both effectiveness and efficiency in terms of training and testing costs. Our comparison was organized around three main research questions, using two datasets. 

From the experiments we find  that models that have been specifically designed for short text
 matching, like MV-LSTM and DRMMTKS, are among the best performing models for both datasets. By taking efficiency and accuracy into account at the same time, ARC-I is the preferred model at least for our proprietary data, which is a good representation of real world e-commerce scenario. Moreover, the performance from a state-of-the-art BERT-based model is mediocre, which we attribute to the fact that the text BERT is pre-trained on is very different from the text we have in product search. We also provide insights that help practitioners choose a well performing method for semantic matching in product search. 

In future work, we aim to categorise queries based on their content and user intent to study the behavior of different methods based on query type, and automatically select the methods accordingly. In addition to that, it would be interesting to investigate other reasons that make matching for product search a challenging task for BERT-based models. Finally, adding a controlled experiment is also beneficial to better validate the results.

\section*{Data and Code}
To facilitate reproducibility of the work in this paper, all codes and parameters are shared at \url{https://github.com/arezooSarvi/sigir2020-eComWorkshop-LTM-for-product-search}

\begin{acks}
This research was supported by Ahold Delhaize. We would also like to thank the reviewers for their thoughtful comments and efforts towards improving our work.
\end{acks}

\bibliographystyle{ACM-Reference-Format}
\bibliography{bibliography}

\end{document}